\documentclass[aip,jcp,twocolumn,10pt]{revtex4-1}
\usepackage{graphicx}%
\usepackage{dcolumn}
\usepackage{amsmath}   
\usepackage[usenames,dvipsnames]{pstricks}
\usepackage{pst-grad} 
\usepackage{pst-plot} 

\usepackage{bm}
\usepackage{longtable}

\begin{document}
\title{Dipolar response of hydrated proteins}
\author{Dmitry V.\ Matyushov}\email{dmitrym@asu.edu}
\affiliation{Center for Biological Physics, Arizona State University,
  PO Box 871504, Tempe, AZ 85287-1504 }

\begin{abstract}
  The paper presents an analytical theory and numerical simulations of
  the dipolar response of hydrated proteins. The effective dielectric
  constant of the solvated protein, representing the average dipole
  moment induced at the protein by a uniform external field, shows a
  remarkable variation among the proteins studied by numerical
  simulations. It changes from 0.5 for ubiquitin to 640 for cytochrome
  \textit{c}. The former value implies a negative dipolar
  susceptibility of ubiquitin, that is a dia-electric dipolar response
  and negative dielectrophoresis. It means that a protein carrying an
  average dipole of $\simeq 240$ D is expected to repel from the
  region of a stronger electric field. This outcome is the result of a
  negative cross-correlation between the protein and water dipoles,
  compensating for the positive variance of the protein dipole in the
  overall dipolar susceptibility. This phenomenon can therefore be
  characterized as overscreening of protein's dipole by the hydration
  shell. In contrast to the neutral ubiquitin, charged proteins
  studied here show para-electric dipolar response and positive
  dielectrophoresis. The protein-water dipolar cross-correlations are
  long-ranged, extending approximately 2 nm from the protein surface
  into the bulk. A similar correlation length of about 1 nm is seen
  for the electrostatic potential produced by the hydration water
  inside the protein. The analysis of numerical simulations suggests
  that the polarization of the protein-water interface is strongly
  affected by the distribution of the protein surface charge.  This
  component of the protein dipolar response gains in importance for
  high frequencies, above the protein Debye peak, when the response of
  the protein dipole becomes dynamically arrested. The interface
  response found in simulations suggests a possibility of a positive
  increment of the high-frequency dielectric constant of the solution
  compared to the dielectric constant of the solvent. This analysis
  provides a theoretical foundation for experimentally observed
  positive increments of the absorption of THz radiation by protein
  solutions.
\end{abstract}
\maketitle

\section{Introduction}
\label{sec:1}
\begin{figure}
  \centering
  \includegraphics*[width=7cm]{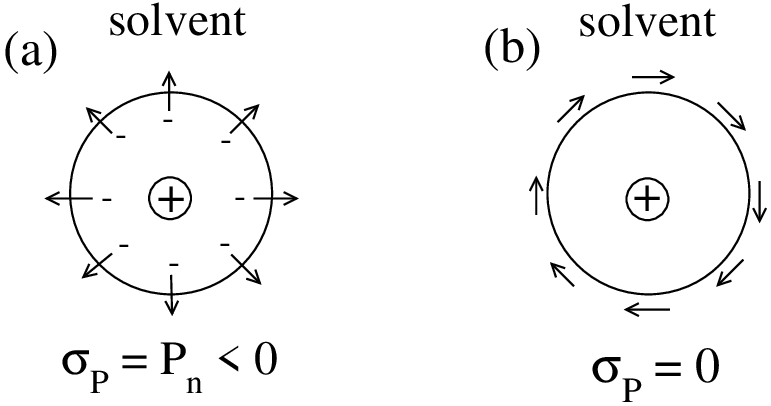}
  \caption{Cartoon of two extreme polarization patterns of a polar
    solvent at the surface of a spherical solute with a positive
    charge at the center.  In panel (a), the solvent dipoles align
    with the electric field of the solute charge. This alignment
    results in surface charge density $\sigma_P=P_n<0$. In panel (b),
    the surface dipoles preserve their preferential in-plane
    orientations characteristic of a free surface of a polar
    liquid.\cite{Sokhan:97} This orientational pattern produces no
    surface charge, $\sigma_P = 0$. }
  \label{fig:1}
\end{figure}

Polarization of the interface is an important component of the
response of a polar substance to an external field. The standard
approach of Maxwell's electrostatics assumes that the interface cuts
through the polarized dipoles of the dielectric, leaving their
monopoles at the surface (Fig.\ \ref{fig:1}a). The density of these
monopoles is the surface charge density $\sigma_P$.  It is given by
the projection $P_n$ of the dipolar polarization vector $\mathbf{P}$
on the outward normal $\mathbf{\hat n}$ to a continuous dielectric
medium.\cite{Landau8,Frohlich} This surface charge is opposite in sign
to an external charge and so the field of the surface charges
compensates the external field. The sum of the two fields makes the
Maxwell field inside the dielectric, which is lower in intensity than
the external field.

\begin{figure}
  \centering
  \includegraphics*[width=5cm]{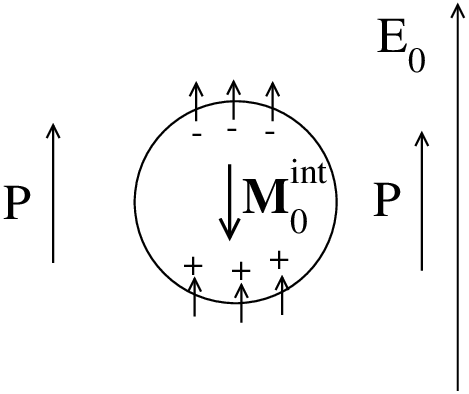}
  \caption{Cartoon of the interface polarization of a spherical void
    in a uniformly polarized polar liquid. The surface charges at the
    interface produce negative and positive lobes of the overall surface
    charge density responsible for the interface dipole
    $M_{0z}^{\text{int}}=\chi_1\Omega_0 E_0$.  When the surface charge
    density disappears because of the in-plane alignment of the
    surface dipoles, the dipole subtracted from the solution is simply
    the product of the solute volume and the uniform polarization of
    the bulk [Eq.\ (\ref{eq:50})]. This is the Lorentz scenario of the
    interface polarization corresponding to $\chi_1=0$. In contrast,
    the Maxwell scenario anticipates a non-zero surface charge
    density, which reduces the dipole taken from the solution from the
    Lorentz value $-P_z\Omega_0$ to the Maxwell value given by Eq.\
    (\ref{eq:39}). This scenario also anticipates a nonzero and
    negative dipolar interface susceptibility $\chi_1$ given by Eq.\
    (\ref{eq:24}). }
  \label{fig:2}
\end{figure}

The same basic considerations apply to the problem of solutions
polarized by a uniform external field $\mathbf{E}_0$. The interface is
now a closed surface enveloping each solute. The polarization field,
uniform in the bulk, becomes inhomogeneous close to the solute-solvent
interface. It generates positive and negative lobes of the surface
charge density (Fig.\ \ref{fig:2}) integrating into an overall
interface (subscript ``int'') dipole $\mathbf{M}_0^{\text{int}}$. Its
calculation is generally a complex problem involving both the effects
of the solute shape and the alteration of the liquid structure by the
solute surface multipoles. A closed-form solution is, however,
possible in the framework of standard dielectric theories for a
spherical void in a dielectric,\cite{Landau8} when all specifics of
the solute-solvent interactions are neglected
\begin{equation}
  \label{eq:39}
  \mathbf{M}_0^{\text{int}} =-3\Omega_0\mathbf{P}/(2\epsilon_s+1) .
\end{equation}
Here, $\epsilon_s$ is the solvent dielectric constant, $\Omega_0$ is
the solute volume, and subscript ``0'' is assigned throughout below to
the solute parameters.  The orientation of the interface dipole is
opposite to the uniform polarization of the medium $\mathbf{P}$.

The appearance of $\mathbf{M}_0^{\text{int}}$ is an interfacial, but
not necessarily a surface phenomenon. This interface dipole is the
integral effect of the inhomogeneous polarization surrounding an
excluded volume of the solute. This polarization perturbation in fact
propagates quite far into the bulk, as we show below, and can be taken
fully into account only in the thermodynamic limit for the solution,
which we represent below as the $k\rightarrow 0$ limit in the inverted
Fourier space of wavevectors $\mathbf{k}$. The surface charge density
$\sigma_P$ is just a convenient mathematical representation of this
highly non-local physical reality in terms of properties assigned to
an infinitely thin mathematical surface enveloping the solute.

The interface dipole arising from a void in a uniformly polarized
liquid will in turn polarize the surrounding solvent. As a result,
the dipole moment of a uniformly polarized solution is given as
\begin{equation}
  \label{eq:50}
  \mathbf{M}_{s} = \mathbf{M}^{\text{liq}} - \Omega_0 \mathbf{P} -
  (2/3)(\epsilon_s-1)\mathbf{M}_{0}^{\text{int}} . 
\end{equation}
The first summand in this equation is the dipole moment of a uniformly
polarized homogeneous liquid. The second term is the dipole moment
reduced from the liquid by putting a solute of volume $\Omega_0$ into
it. Finally, the last summand is the polarization of the liquid by the
interface dipole.

The interface dipole $\mathbf{M}_0^{\text{int}}$ exists at any closed
interface in a polarized medium, even in the absence of solute's own
charges. When calculated according to Eqs.\ (\ref{eq:39}) and
\eqref{eq:50}, it will lower the dielectric constant of the solution
$\epsilon$ compared to the dielectric constant $\epsilon_s$ of the
homogeneous liquid. In contrast, if the solute possesses its own
dipole, it will align along the external field $E_0$ producing an
average permanent dipole $\langle \mathbf{M}_0\rangle_E$. This dipole
moment will enhance the dielectric response of the solution.  The
overall dipole moment associated with a solute will be the sum of the
intrinsic permanent dipole and the dipole induced at the dielectric
interface\cite{DMpre:10}
\begin{equation}
  \label{eq:12}
 \mathbf{M}_0 = \langle \mathbf{M}_0\rangle_E + \mathbf{M}_0^{\text{int}}  ,
\end{equation}
where $\langle \dots \rangle_E$ refers to the statistical average in
the presence of the field.

Both the permanent and interface components of $\mathbf{M}_0$ depend
on the ability of the interface to polarize. The very basic physics
outlined in Fig.\ \ref{fig:1}a assumes the dipoles of the medium to
have the ability to freely change their orientations in order to align
along an external electric field. While this is probably the case for
solvation of small ions in polar liquids, solvation of larger solutes
might present a challenge to this picture.

Dipoles of polar liquids preferentially orient in-plane at
interfaces.\cite{Lee:84,Sokhan:97,Bratko:09,DMepl:08,Rossky:10,DMjcp2:11} Unless
an external field rotates the dipoles off-plane, such orientational
structure eliminates the surface charge since $\sigma_P = P_n \simeq
0$ in this case (Fig.\ \ref{fig:1}b). The standard boundary conditions
of continuum electrostatics do not apply to this scenario which
implies
\begin{equation}
  \label{eq:41}
  \mathbf{M}_0^{\text{int}} = 0, 
\end{equation}
instead of the standard electrostatic result listed in Eq.\
(\ref{eq:39}). 

Equations (\ref{eq:50}) and (\ref{eq:41}) suggest that carving a void
from a dielectric removes the dipole moment equal to the product of
the uniform bulk polarization and the void volume. Such a solution
would appear in the standard theories of dielectrics\cite{Scaife:98}
if the void in the dielectric had the interface of a Lorentz virtual
cavity, which has no surface polarization by definition. We will dubb
this outcome, corresponding to $\sigma_P=0$ in Fig.\ \ref{fig:1}b, as
the ``Lorentz scenario''. On the other hand, when
$\mathbf{M}_0^{\text{int}}$ from Eq.\ (\ref{eq:39}) is substituted
into Eq.\ (\ref{eq:50}), the two last terms combine into the same
$\mathbf{M}_0^{\text{int}}$, which becomes the dipole subtracted from
the homogeneous liquid upon insertion of the solute.  The rules of
electrostatics therefore predict that the dipole removed from the
solution will be, at $\epsilon_s\gg 1$, just a small fraction of the
Lorentz dipole. Since this scenario follows from the standard
electrostatics with Maxwell's boundary conditions at the dielectric
interface, we will call this outcome, and the corresponding interface
polarization, the ``Maxwell scenario'' (Figs.\ \ref{fig:1}a and
\ref{fig:2}).

Water presents a particularly important study case for the Lorentz
scenario of Eq.\ (\ref{eq:41}). Large solutes, over 1 nm in
size,\cite{ChandlerNature:05,Ball:08} break the network of hydrogen
bonds of bulk water, resulting in preferential in-plane orientation of
the interfacial water
dipoles.\cite{Lee:84,Sokhan:97,Bratko:09,Rossky:10} This phenomenon,
general for polar fluids,\cite{DMepl:08,DMpre1:08} is further
amplified for water interfaces by the high energy of water's hydrogen
bonds.\cite{DMjcp2:11}

The orientational structure of interfacial water has macroscopically
observable consequences, both mechanical and electrostatic.
Mechanical consequences include the rotation of hydrated nanometer
solutes by external fields \cite{Daub:09} and slipping of the boundary
layers in the hydrodynamic flow.\cite{Barrat:99} For electrostatic
observables, the electric field inside cavities formed in uniformly
polarized dipolar liquids\cite{DMepl:08} and in water\cite{DMjcp2:11}
seem to follow the scenario of an unpolarized interface sketches in
Fig.\ \ref{fig:1}b. One can therefore anticipate that in-plane dipolar
orientations might be preserved, at least in patches, if the external
field is not sufficiently strong to compete with interface hydrogen
bonds forcing water dipoles orient in-plane. If this is the case, the
boundary conditions of the dielectric response problem will alter,
thus affecting all relevant polar response functions.

Electrostatic interactions are critical for biological
function.\cite{Tessier:03,Gitlin:06,Warshel:06} Most biomolecules and
all cellular membranes carry charges.\cite{Sear:03} Electrostatic
solvation and interactions affect the stability of folded proteins and
their aggregation and crystallization.
\cite{Richardson:02,Lawrence:07,Pace:09} Therefore, the question of
what is the dipolar polarization at the interface of a hydrated
biomolecule is critical for structural biology and
bioenergetics.\cite{DMpccp:10}

Surfaces of proteins, and of all biomaterials in general, are
obviously chemically and electrostatically
heterogeneous.\cite{Giovambattista:08} One therefore cannot expect a
clear-cut scenario of either in-pane dipoles or dipoles fully aligned
along the electric field. This study in fact shows that none of the
scenarios sketched in Fig.\ \ref{fig:1} presents a complete
description of hydrated proteins, which explore a much wider range of
possibilities allowing them to tune their response to global and local
\textit{in vitro} fields. To grasp this complexity, we ask what would
be a minimal set of coarse-grained parameters describing the
water-protein interface.  We approach this question here by first
presenting an analytical theory framing the problem in terms of a set
of interface susceptibilities, followed by numerical simulations of
several hydrated globular proteins. The property of interest is the
average dipole moment $M_0$ [Eq.\ \eqref{eq:12}] induced at the
protein by an external electric field. As such, this is a fundamental
and well-defined physical quantity related to broad-band dielectric
spectroscopy of solutions, not considered here, and directly probed by
dielectrophoresis of protein
solutions\cite{Washizu:94,JonesBook:95,HughesBook:03} discussed below.

\section{Dielectric considerations}
\label{sec:2}
According to the separation of the solute dipole into the intrinsic
permanent and interfacial components, one can define linear dipolar
susceptibilities for the corresponding dipoles along the external
field.  If the $z$-axis of the laboratory frame is set along the
external field, one gets for the dipolar (``d'') and interfacial
(``int'') parts of the response
\begin{equation}
  \label{eq:40}
  \begin{split}
    &\chi_{0}^d  = \langle M_{0z}\rangle_E /(\Omega_0 E_0),\\
    &\chi_{0}^{\text{int}} = M_{0z}^{\text{int}}/ (\Omega_0 E_0) .
  \end{split}
\end{equation}

We first focus on the solute permanent dipole and consider the
first-order perturbation expansion for the corresponding
susceptibility\cite{Frohlich}
\begin{equation}
  \label{eq:1}
  \chi_0^d=(\beta/3\Omega_0) \langle \delta \mathbf{M}_0 \cdot \delta
  \mathbf{M}\rangle ,
\end{equation}
where $\beta=1/(k_{\text{B}}T)$ is the inverse temperature.  Further,
$\delta \mathbf{M}_0 = \mathbf{M}_0 - \langle \mathbf{M}_0\rangle $
and $\delta \mathbf{M} = \mathbf{M} - \langle\mathbf{ M}\rangle $ are
the deviations of the solute ($\mathbf{M}_0$) and total sample
($\mathbf{M}$) dipole moments from their corresponding average values;
$\langle \dots \rangle$ refers to a statistical average in the absence
of an external field. Since the solute can in principle be charged,
keeping the variation $\delta\mathbf{M}_0$ eliminates the dependence
of its dipole on the origin of the system of coordinates. On the other
hand, $\delta\mathbf{M}$ can be replaced with $\mathbf{M}$ if the
sample is neutral and isotropic.

The dipolar susceptibility
\begin{equation}
  \label{eq:43}
  \chi_0 = \chi_0^d + \chi_0^{\text{int}}   
\end{equation}
connects average dipole of the solute to a weak external field $E_0$
which varies on a length-scale significantly larger than the solute
dimension. This electric field creates an excess chemical potential of
the solute\cite{Landau8}
\begin{equation}
  \label{eq:5}
  \Delta \mu_0 = - \frac{1}{2}  M_{0z}\ E_0 ,
\end{equation}
where $\mathbf{M}_0$ is given by Eq.\ \eqref{eq:12}. The dipolar
susceptibility therefore follows from the derivative of $\Delta \mu_0$
over the external field strength: $\chi_0\Omega_0 E_0 = - \partial
\Delta \mu_0 / \partial E_0$. Since the external field $E_0$ is well
defined by the charge density at the plates of a planar capacitor in
the dielectric experiment or by the light intensity in absorption
measurements, the corresponding susceptibility is a well defined
parameter as well.

This susceptibility is obviously distinct from the susceptibility of
the protein solution responding to the macroscopic Maxwell field $E$.
One can consider the average solute dipole created in response to
$E$. In that case, one needs a connection between the two fields. This
connection, $E=E_0/ \epsilon$ (where $\epsilon$ is the solution
dielectric constant), is again straightforward in dielectric
measurements, but depends on the simulation protocol in numerical
simulations.\cite{Neumann:86,King:91} It simplifies, however,
significantly for tin-foil implementation of the Ewald sums
representing Coulomb interactions in simulations with periodically
replicated simulation cell.\cite{Neumann:86} In that case, which is
followed in this study, $E=E_0$ and $\chi_0^d$ from MD trajectories
gives the response to the Maxwell field $E$. A more elaborate theory,
which we present below, is, however, needed to obtain the interface
susceptibility $\chi_0^{\text{int}}$. Once this is done, one can
follow the standard convention to introduce the statistical (that is
obtained from the variance of the dipole moment) dielectric
constant\cite{Warshel:06} of the protein
\begin{equation}
  \label{eq:2}
  \epsilon_0 =1 + 4\pi \chi_0 .
\end{equation}

The dielectric constant $\epsilon_0$, used here to quantify the solute
dipolar response, is neither the dielectric constant of the solution
$\epsilon$ reported by the dielectric experiment
\cite{Takashima:89,Pethig:92,Oleinikova:04} nor it is the screening
parameter used to describe Coulomb interactions between charges inside
the solute, such as atomic charges of protein
residues.\cite{Warshel:06,Vicatos:09,Isom:10} We also stress that
$\epsilon_0$ defined by Eqs.\ \eqref{eq:43} and \eqref{eq:2}
represents the dipolar response of the entire protein (irrespective of
its shape) and tells nothing about dielectric properties of any region
inside the protein.\cite{King:91} The definition of the solute dipolar
response in terms of the external field $E_0$, instead of the Maxwell
field $E$, is convenient for a number of reasons, including the fact
that one does not need to calculate the solution dielectric constant
$\epsilon$, which requires additional modeling.

The susceptibility
\begin{equation}
  \label{eq:25}
  \chi_0^d = \chi_{00}+\chi_{0s} 
\end{equation}
is a sum of the direct solute component $\chi_{00} \propto \langle
\delta M_0^2 \rangle $ and a cross-correlation term $\chi_{0s} \propto
\langle \delta \mathbf{M}_0 \cdot \delta \mathbf{M}_s\rangle $, where
$\mathbf{M}_s = \mathbf{M} -\mathbf{M}_0$ is the solvent dipole
moment. While $\chi_{00}$, the dielectric susceptibility defined for a
finite solute volume $\Omega_0$,\cite{King:91} is obviously positive,
one wonders what is the sign and relative magnitude of the $\chi_{0s}$
component.\cite{Loffler:97} The answer depends strongly on the details
of the dipolar polarization of the solute-solvent interface.

One needs to recognize that the correlator of the solute dipole
$\mathbf{M}_0$ with the entire sample dipole $\mathbf{M}$ in Eq.\
\eqref{eq:1}, instead of the self-correlator $ \langle \delta M_0^2
\rangle $ typically appearing in macroscopic theories of dielectrics,
substitutes for the boundary conditions used in these theories. The
boundary conditions represent the physical fact that any treatment of
the dipolar polarization of a finite sample should include surface
charges if the sample is placed in vacuum or the polarization of the
surrounding medium if the finite sample is a part of an infinite
dielectric material.\cite{Frohlich} This notion also implies that
understanding and potentially modeling of the cross-correlation
susceptibility $\chi_{0s} \propto \langle \delta \mathbf{M}_0 \cdot
\delta \mathbf{M}_s\rangle$ allows one to substitute the boundary
conditions of standard electrostatics, which rely on the bulk
dielectric constant,\cite{Landau8,Frohlich} with microscopic rules
incorporating various polarization scenarios, such as two extremes
sketched in Fig.\ \ref{fig:1}.  This perspective is particularly
important for studies of solvation at the nano-meter scale. The number
of first-shell waters of a typical globular protein reaches the
magnitude of $N_1\simeq 200-500$, separating them, and potentially
other nearby shells, into a sub-ensemble. The properties of this
sub-ensemble might dramatically differ from those of the bulk
solvent. The language of interfacial susceptibilities might better
grasp this reality than bulk properties typically used in standard
theories to construct the interfacial response functions.

Accordingly, we introduce below the dipolar susceptibility of the
surface charge density $\chi_1$ which will incorporate all possible
boundary conditions at the surface of the solute and will reproduce
the Lorentz and Maxwell scenarios as special cases. We will use this
susceptibility to both provide a connection between $\chi_{0s}$
and $\chi_{00}$ components of the dipolar response function $\chi_0^d$
and derive a relation for $\chi_0^{\text{int}}$. This formalism will
allow us to analyze the results of numerical simulations of protein
solutions presented next.

\section{Response functions}
\label{sec:3}
A general insight into the understanding of the solvent polarization
surrounding a solute can be gained from the inhomogeneous response
function of the dipolar polarization in the solute's vicinity. These
types of problems typically involve integral convolutions in real
space, which become algebraic relations in inverted
$\mathbf{k}$-space. In the presence of the solute, the dipolar
response function $\bm{\chi}(\mathbf{k}_1,\mathbf{k}_2)$, which is a
rank-two tensor, becomes a function of two wave-vectors,
$\mathbf{k}_{1}$ and $\mathbf{k}_{2}$.\cite{DMjcp1:04} This is a
reflection of the inhomogeneous character of the problem, in contrast
to the response function of the homogeneous solvent
$\bm{\chi}_s(\mathbf{k})$, which depends on one wave-vector only.

The modification introduced to the solvent response by the solute can
be given as an inhomogeneous correction to $\bm{\chi}_s(\mathbf{k})$
as follows
\begin{equation}
  \label{eq:10}
  \bm{\chi}(\mathbf{k}_1,\mathbf{k}_2)=
  \bm{\chi}_s(\mathbf{k}_1)\delta_{\mathbf{k}_1,\mathbf{k}_2} - (\chi_1/ \chi_1^{\text{M}})
   \bm{\chi}_0(\mathbf{k}_1,\mathbf{k}_2) .
\end{equation}
Here, the Kronecker delta $\delta_{\mathbf{k}_1,\mathbf{k}_2}$ is
normalized to the sample volume $V$,
$\delta_{\mathbf{0},\mathbf{0}}=V$.  Further, the homogeneous liquid
has isotropic symmetry, which is broken in $\mathbf{k}$-space by the
wave-vector $\mathbf{\hat k}=\mathbf{k}/k$ introducing axial symmetry
to the problem. The second-rank tensor functions are then fully
described by two scalar projections, longitudinal (L) and transverse
(T), onto two diadic tensors,\cite{Wertheim:71}
$\mathbf{J}^L=\mathbf{\hat k}\mathbf{\hat k}$ and
$\mathbf{J}^T=\mathbf{1}-\mathbf{\hat k}\mathbf{\hat
  k}$. Correspondingly, the response function of the homogeneous
solvent is given as the sum of the longitudinal and transverse
components,\cite{DMjcp1:04} $\bm{\chi}_s(\mathbf{k}) =
\chi_s^L(k)\mathbf{J}^L + \chi_s^T(k)\mathbf{J}^T$.  Their $k=0$
values are bulk susceptibilities connected to the liquid dielectric
constant $\epsilon_s$
\begin{equation}
  \label{eq:29}
  \begin{split}
     \chi_s^L(0)=&(\epsilon_s-1)/(4\pi\epsilon_s),\\
     \chi_s^T(0)=&(\epsilon_s-1)/(4\pi).
  \end{split}
\end{equation}

The parameter $\chi_1$ in front of heterogeneous part of the response
$ \bm{\chi}_0(\mathbf{k}_1,\mathbf{k}_2)$ in Eq.\ \eqref{eq:10} is the
dipolar susceptibility of the solute interface coarse-grained into a
spherical surface. It appears from the expansion of the surface charge
density induced by a uniform external field in Legendre polynomials of
the polar angle $\theta$ between a radius-vector at the surface and
the external field, $\sigma_P(\theta)=\sum_{\ell} \sigma_{\ell}
P_{\ell}(\cos \theta)$.  Only the first-order, dipolar component
$\ell=1$ contributes to the dipolar polarization field
$\mathbf{P}(\mathbf{r})$ of the liquid. The susceptibility $\chi_1$
connects $\sigma_1$ to the field strength
\begin{equation}
  \label{eq:44}
     \sigma_1 = \chi_1  E_0 .   
\end{equation}
Correspondingly, the interface dipole is calculated by multiplying
$\sigma_P(\theta)$ with the surface radius-vector and integrating over
the closed surface. The result is
\begin{equation}
  \label{eq:52}
     M_{0z}^{\text{int}} = \chi_1\Omega_0 E_0 
\end{equation}
and, from Eq.\ (\ref{eq:40}), $\chi_0^{\text{int}}=\chi_1$. 

The derivation of $\bm{\chi}(\mathbf{k}_1,\mathbf{k}_2)$ for a void in
a polar liquid that we discuss in Appendix \ref{appA} yields the
standard Maxwell result for $M_0^{\text{int}}$ in Eq.\ (\ref{eq:39}) and a
negative $\chi_1$
\begin{equation}
  \label{eq:24}
  \chi_1^{\text{M}} = -\frac{3}{2\epsilon_s+1}\ \chi_s^L(0).
\end{equation}
In contrast, the Lorentz scenario corresponds to $\chi_1=0$.  Since we
do not want to be limited by the solution describing a void and
instead want to introduce an interfacial polarization induced by the solute,
$\chi_1$ in Eq.\ (\ref{eq:10}) is an unspecified susceptibility
calculated below from numerical simulations.

The polarization of the solvent in response to a field
$\mathbf{E}_0(\mathbf{r})$, which might include both solute and
external charges, is given by the convolution in inverted space
\begin{equation}
  \label{eq:14}
  \mathbf{\tilde P}(\mathbf{k}) =
  \bm{\chi}(\mathbf{k},\mathbf{k}')*\mathbf{\tilde E}_0(\mathbf{k}'), 
\end{equation}
where tildes over vectors specify inverted-space fields and asterisks
denotes both the tensor contraction and $\mathbf{k}$-space
integration, such as
\begin{equation}
  \label{eq:15}
  \mathbf{\tilde A}*\mathbf{\tilde B}= \int \mathbf{\tilde A}\cdot
  \mathbf{\tilde B}\ d\mathbf{k}/(2\pi)^3 .
\end{equation}

We now want to apply the inhomogeneous response function in Eq.\
\eqref{eq:10} to calculate the cross-correlation term $\chi_{0s}$ in
Eqs.\ \eqref{eq:40} and (\ref{eq:25}).  In order to approach this
problem, we will calculate the solvent dipole $\mathbf{M}_s$ induced
by the inhomogeneous field of an instantaneous solute dipole
$\mathbf{M}_0$. We will assume a certain separation of time-scales to
perform this calculation. Specifically, the solvent is assumed to be
much faster than the solute and thus able to follow adiabatically
every instantaneous configuration of the solute electric field. This
approximation is typically correct for hydrated proteins since
high-frequency protein vibrations produce a relatively minor effect on
the protein field sensed by hydration water.\cite{DMpre:11}

For this problem, the external electric field now becomes
\begin{equation}
  \label{eq:17}
  \mathbf{\tilde E}_0(\mathbf{k}) = \mathbf{\tilde T}(\mathbf{k})\cdot
  \mathbf{M}_0, 
\end{equation}
where
\begin{equation}
  \label{eq:18}
  \mathbf{\tilde T}(\mathbf{k}) = - 4\pi\frac{j_1(kR)}{kR}
  \mathbf{\tilde D} , 
\end{equation}
is the $\mathbf{k}$-space dipolar tensor of a spherical solute with an
effective radius $R$; $\mathbf{\tilde D}=3\mathbf{\hat k}\mathbf{\hat
  k} -\mathbf{1} = 2\mathbf{J}^L-\mathbf{J}^T$ and $j_{\ell}(x)$ is
the spherical Bessel function.

The overall solvent dipole $\mathbf{M}_s$ can be obtained by
integrating the dipolar polarization field $\mathbf{P}(\mathbf{r})$
over the volume occupied by the solvent, $\Omega=V-\Omega_0$. This
direct-space integration is equivalent, in inverted space, to
taking $k=0$ limit for the polarization of the entire sample
($V\rightarrow \infty$ in the thermodynamic limit) and subtracting the
convolution of the polarization field with the step function
$\theta_0(\mathbf{r})$, equal to unity within the solute and zero
everywhere else.\cite{DMjcp1:04} One then gets for the inverted-space
fields
\begin{equation}
  \label{eq:23}
  \mathbf{M}_s = \mathbf{\tilde P}(0) - 
                 \mathbf{\tilde P}(\mathbf{k})*\tilde\theta_0(\mathbf{k}) ,
\end{equation}
where $\mathbf{\tilde P}(\mathbf{k})$ is given by Eq.\ (\ref{eq:14})
and 
\begin{equation}
  \label{eq:20}
  \tilde\theta_0(\mathbf{k}) =\int_{\Omega_0} e^{i\mathbf{k}\cdot\mathbf{r}}
  d\mathbf{r} . 
\end{equation}

\begin{figure}
  \centering
  \includegraphics*[width=7cm]{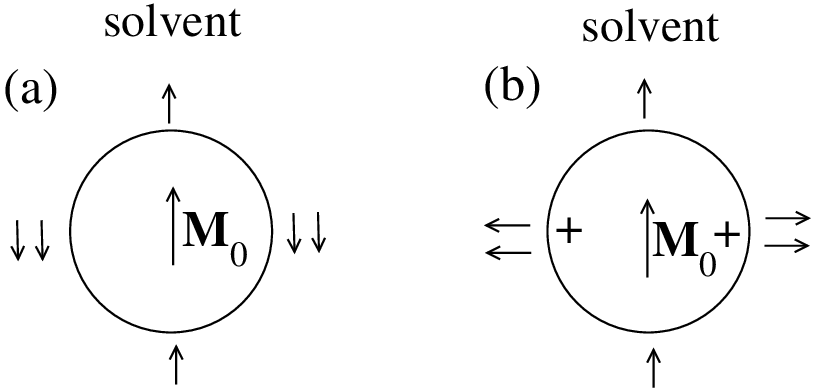}
  \caption{Cartoon of the solvent polarization around a dipolar
    solute. A higher number of solvent dipoles in the equatorial plane
    of the solute dipole $\mathbf{M}_0$ (a) yields a negative value of
    the response function $\chi_{0s}$.  In panel (b), the addition of
    equatorial charges to the solute does not alter the solute dipole
    calculated relative to the center of the sphere, but polarizes
    equatorial waters in a direction orthogonal to $\mathbf{M}_0$. The
    results is a positive $\chi_{0s}$, as indeed found for all charged
    proteins here. }
  \label{fig:3}
\end{figure}

We show in Appendix \ref{appA} that the second summand in Eq.\ (\ref{eq:23})
is zero when the continuum, $k\rightarrow 0$ limit is taken in the
homogeneous solvent response function $\bm{\chi}_s(\mathbf{k})$. We
will use this limit throughout below since the typical size of the
protein $2R$ significantly exceeds the diameter of water.  Within this
approximation one arrives (see Appendix \ref{appA}) at $\chi_{0s}$ and then
at the following connection between the $\chi_{0s}$ and $\chi_{00}$
response functions in terms of the interfacial dipolar susceptibility
$\chi_1$
\begin{equation}
  \label{eq:21}
  \chi_{0s} = -\frac{2}{3}\frac{\epsilon_s-1}{\epsilon_s} 
   \left( 1 - (4\pi/3) \epsilon_s \chi_1 \right) \chi_{00}. 
\end{equation}
From this equation and Eq.\ (\ref{eq:25}), one additionally have
\begin{equation}
  \label{eq:54}
  \frac{\chi_0^d}{\chi_{00}}= \frac{\epsilon_s+2}{3\epsilon_s} +
  \frac{8\pi}{9}(\epsilon_s-1)\chi_1 .
\end{equation}

If the term in the brackets in Eq.\ (\ref{eq:21}) is positive, as is
the case in both the Lorentz and Maxwell scenarios, $\chi_{0s}$ is
negative. Figure \ref{fig:3}a illustrates the physical origin of this
result.  The solvent polarization is a sum of two compensating
contributions. The solvent dipoles at the poles of the solute dipole
will predominantly orient parallel to $\mathbf{M}_0$, while equatorial
solvent dipoles will orient antiparallel to $\mathbf{M}_0$. Since
there are more equatorial dipoles than there are pole dipoles,
$\chi_{0s}<0$ for the overall solvent polarization.

It is instructive to see what are the numerical outputs for $\chi_0^d $
in the Maxwell and Lorentz scenarios sketched in Fig.\ \ref{fig:1}. In
the Lorentz case of $\chi_1=0$ one gets
\begin{equation}
  \label{eq:3}
  \chi_0^d = \frac{\epsilon_s+2}{3\epsilon_s} \chi_{00} . 
\end{equation}
The correction term in front of $\chi_{00}$ is the Lorentz cavity
field which indeed appears in polar response when the surface of a
cavity cut from the dielectric is not
polarized.\cite{Frohlich,Scaife:98}

When the Maxwell result [Eq.\ (\ref{eq:24})] for the susceptibility
$\chi_1$ is used in Eq.\ (\ref{eq:21}) one gets
\begin{equation}
  \label{eq:4}
  \chi_0^d = \frac{3}{2\epsilon_s+1} \chi_{00} .
\end{equation}
This correction factor is the well-known cavity field of the theories
of dielectrics.\cite{Frohlich,Scaife:98} These arguments stress again
(see above) that specifying an algorithm of calculating $\mathbf{M}_s$
in terms of $\mathbf{M}_0$ leads to a route to formulate a theory of
polar response, with the prescription of Figs.\ \ref{fig:1}a and
\ref{fig:2} corresponding to Maxwell's
electrostatics.\cite{Landau8,Frohlich,Scaife:98}

Equations (\ref{eq:3}) and (\ref{eq:4}) are special cases of a more
general result
\begin{equation}
  \label{eq:53}
  \frac{\chi_0^d}{\chi_{00}} = \frac{E_c}{E_0}
\end{equation}
connecting the cavity field inside the solute $E_c$ with the ratio of
two response functions. The notion ``cavity'' here implies that this
field is produced by the solvent polarized by the external field and
does not include any reaction fields of the solute charges. While the
two contributions into the overall electric field inside the solute
might seem to be hopelessly entangled, they are in fact separable in
the frequency domain. The solute and corresponding reaction fields are
dynamically frozen at frequencies above the Debye peak of the solute,
in most practical cases well below the Debye peak of the water
component of the solution. The observation of the dielectric increment
of the water's Debye peak as a function of the solute concentration in
the limit of ideal solution provides a direct access to the cavity
field\cite{DMjcp2:11} and, by Eqs.\ (\ref{eq:54}) and (\ref{eq:53}),
to the ratio of the two susceptibilities and $\chi_1$. The relation of
these susceptibilities to dielectrophoresis discussed below is yet
another connection to the laboratory experiment.

A significant difference in the solute dipolar response predicted by
Lorentz and Maxwell polarization scenarios offers an opportunity to
control the interaction of the solute dipole with an external field by
changing the distribution of the surface charge. For example, consider
the effect of surface charges placed in the equatorial region of the
global solute dipole in Fig.\ \ref{fig:3}b. These charges will not
change the overall solute dipole calculated relative to the sphere's
center, but will orient solvent dipoles perpendicular to the global
dipolar field of the solute and can potentially invert the sign of
$\chi_{0s}$. This is indeed what we find in our simulations of charged
hydrated proteins.

One can also anticipate some combination of the solute shape and
charge distribution that will produce a negative net result for
$\chi_{0}$, when negative $\chi_{0s}$ exceeds in magnitude
$\chi_{00}$. This outcome, which can be characterized as
overscreening\cite{Ballenegger:05} of the solute dipole by the
hydration layer, would correspond to a dia-electric response of the
solute, i.e.\ its repulsion from the region of a more intense electric
field.  We find this result in our simulations of the neutral
ubiquitin (ubiq) protein.

\section{Results of MD Simulations}
\label{sec:4}
Equation (\ref{eq:21}) is a central result of our derivation. It
connects the two component of the solute dipolar susceptibility and
the interface moment $M_0^{\text{ind}}$ to one single property, the
dipolar susceptibility of the interface $\chi_1$.  This susceptibility
is a coarse-grained parameter incorporating the effects of both the
shape and surface charge distribution of the solute.  Since both
$\chi_{0s}$ and $\chi_{00}$ are in principle accessible from numerical
simulations, Eq.\ \eqref{eq:21} allows us to construct the overall
dipolar response of a hydrated solute without additional assumptions
on the nature of polarization boundary conditions at the interface.

Four globular proteins, reduced form of cytochrome \textit{c} (cytC,
100 ns, PBD database 2B4Z), ubiquitin (ubiq, 172 ns, 1UBQ), lysozyme
(lys, 153 ns, 3FE0), and reduced form of cytochrome B562 (cytB, 123
ns, 256B) were studied by long, 100--172 ns all-atom Molecular
Dynamics (MD) simulations.  The overall length of the simulation
trajectories was $>0.5$ $\mu$s.  All proteins were solvated in large
numbers of TIP3P waters to achieve mM protein concentration typically
used in experimental solution measurements. The number of waters in
the simulation box were $N_s= 33189$ (cytC), 27918 (ubiq), 27673 (lys),
and 33268 (cytB). From these proteins, cytB is the only protein with
the overall negative charge $Z$ (Table \ref{tab:1}), and it was chosen
mostly for that reason.

The importance of long simulations has been recognized in previous
attempts to address the dielectric properties of protein solutions,
\cite{Smith:93,Simonson:96,Boresch:00,Schroder:06,Rudas:06} but they
had never reached the length and system size reported here. In
particular, Steinhauser and co-workers
\cite{Loffler:97,Boresch:00,Rudas:06} have pointed to a significant
contribution of $\chi_{0s}$ susceptibility to the overall dipolar
response of the solution, but could reach only qualitative conclusions
from trajectories available to them ($\simeq 13-30$ ns). Many previous
attempts to simulate protein solutions had suffered from even shorter
trajectories and far smaller simulation boxes. The aim of this round
of simulations is to extend previous simulation studies to reliably
calculate both $\chi_{00}$ and $\chi_{0s}$ components of the solute
dipolar response. We therefore do not approach here the calculation of
the overall dielectric response of the protein solution and instead
focus on the question of how polar, as probed by $\chi_0$, a hydrated
protein can be. The details of the simulation protocol can be found in
the Supplementary Material (SM)\cite{supplJCP} and we proceed directly
to the results.

\begin{table*}
\begin{ruledtabular}
  \centering
  \caption{\label{tab:1} Dielectric parameters of hydrated proteins from MD simulations.}
   \begin{tabular}{lcccccccccc}
    Protein\footnotemark[1] & $Z$ & $R$ (\AA) & $\epsilon_0$ & $\langle M_0 \rangle$ (D) 
             &  $\chi_{0}^d$ \footnotemark[2] & $\chi_{00} $  &
             $\chi_{0s}$ & $\chi_0^{\text{int}}\times 10^3$ & $\chi_1/ \chi_1^{\text{M}}$ 
            & $\kappa_1$ \footnotemark[3]\\
   \hline
    Lys   & $+7$ & 19.2 &  62 &  223 & 4.9 & 3.9 &  1.0 & 3.7 & $-2.8$ &0.3\\
    CytC  & $+6$ & 18.7 & 639 & 367 & 51 & 38 & 13 & 3.9 & $-3.1$ &5.7 \\
    Ubiq   & $0$  & 17.8 & 0.5 & 244 & $-0.04$ & 0.17 & $-0.21$ &  $-2.3$ &
    $1.78$ & 0.2\\
    CytB  & $-5$  & 23.5 &  96 & 196  & 7.5 & 6.1 & 1.4 & 3.6 & $-2.8$ & 21.6 \\
  \end{tabular}\\
\footnotetext[1]{$Z$ is the overall charge of the protein, $R$ is the
  effective radius, the dipole moment $\langle M_0\rangle$ is
  calculated from the protein
  charges relative to the center of mass and averaged over the simulation trajectory.}
\footnotetext[2]{The standard deviation of $\chi_0^d$ calculated according to the standard
procedures explained in more detail in the SM\cite{supplJCP} are: 0.1(2\%) (lys), 1.6(3\%) (cytC),
$6\times 10^{-3}$(20\%) (ubiq), 0.3(2\%) (cytB).}
\footnotetext[3]{Compressibility of the first hydration shell,
  $\kappa_1=\langle(\delta N_1)^2\rangle/ \langle N_1\rangle $, $N_1$
  is the number of waters in the first shell defined as the layer of
  3 \AA\ thickness measured from the protein vdW surface.}
\end{ruledtabular} 
\end{table*}

\subsection{Static properties}
\label{sec:4-1}
The proteins studied here are similar in their effective size, as
calculated from the volume inside the solvent-accessible surface
enveloping the molecule.\cite{Till:10} The average dipole moments
$\langle M_0 \rangle$ calculated from the protein partial charges and
averaged over the trajectories are also close in magnitude and are in
basic agreement with dipole moments typically reported by dielectric
spectroscopy of protein solutions\cite{Takashima:89} (Table
\ref{tab:1}).  The values of the dipole moments were calculated here
relative to the protein center of mass since these dipoles appear in
the equations of motion establishing the torque applied to the protein
by an external electric field.\cite{Rudas:06} While the protein size
and the dipole moment appear to be generic for the set of globular
proteins studied here, the dipolar susceptibility, i.e.\ the variance
of the protein dipole is quite specific for a given protein. We find a
remarkably broad range of dipolar susceptibilities $\chi_0$ and the
corresponding values of $\epsilon_0 = 1 + 4\pi\chi_0$ among the
proteins studied here (Table \ref{tab:1}).

The polarity of the protein, as measured by $\chi_0$ or $\epsilon_0$,
does not seem to correlate with either the magnitude or the sign of
the overall protein charge. In fact, values of $\epsilon_0$ are close
for positively charged lys ($Z=+7$) and negatively charged cytB
($Z=-5$). A significant variation in the values of $\epsilon_0$ found
here (Table \ref{tab:1}) seems to originate from differences in surface
charge distributions of the proteins and the corresponding polarization
of the hydration water.

The cross-correlation susceptibility $\chi_{0s}$ also varies
significantly among the proteins, both in magnitude and sign. We find
$\chi_{0s}$ positive for the charged proteins. This observation can be
explained in terms of the water polarization scenario pictured in
Fig.\ \ref{fig:3}b. In contrast, $\chi_{0s}$ is negative for the
neutral ubiq, in agreement with the dielectric arguments presented
above. However, in contrast to the standard expectations, $\chi_0^d$
is negative. This outcome implies a dia-electric response or negative
dielectrophoresis. The relative error of this claim is 20\% (Table
\ref{tab:1}).  This is because $\chi_0^d$ comes as a small number from
subtraction of two relatively large numbers, $\chi_{00}$ and
$\chi_{0s}$, and converges very slowly even on the longest trajectory
we have run in this study (see the SM\cite{supplJCP}).

The results for the interface susceptibility $\chi_1$ are consistent
with the results for $\chi_{0s}$ (Table \ref{tab:1}). The
susceptibility $\chi_1$ not only exceeds the Maxwell
$\chi_1^{\text{M}}$ [Eq.\ (\ref{eq:24})] in magnitude, but has the
sign opposite to it. A positive $\chi_1$ physically means that the
dipole produced by the surface charge density $\sigma_P$ (Fig.\
\ref{fig:2}) is oriented along the field and not opposite to the field
as the Maxwell scenario would suggest. As for $\chi_{0s}$, the origin
of this outcome should be sought along the lines illustrated in Fig.\
\ref{fig:3}b, which shows that different regions of the interface
polarization can add constructively or destructively depending on the
distribution of the surface charge.

\begin{figure}
  \centering
  \includegraphics*[width=8cm]{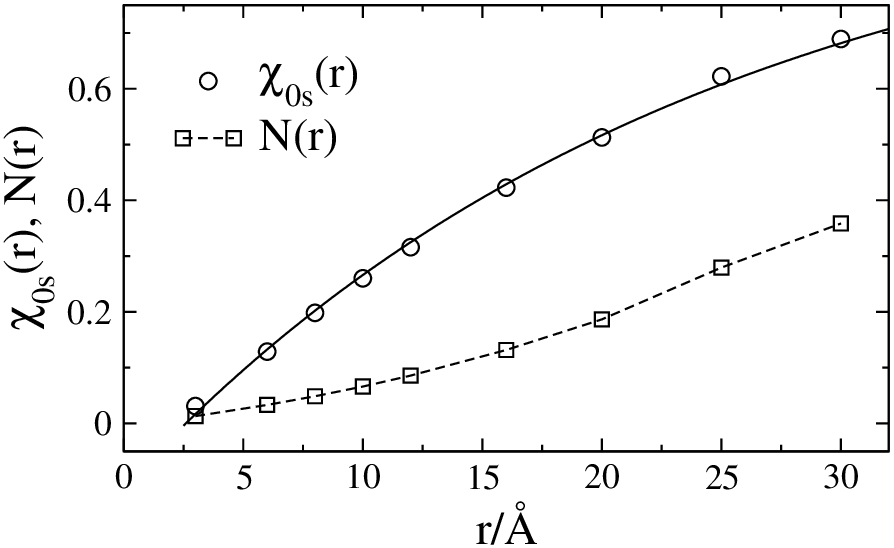}
  \caption{Reduced functions $\chi_{0s}(r) / \chi_{0s}$ (circles) and $N(r) /
    N_s$ (squares) calculated from the layer of water of thickness $r$
    surrounding the vdW surface of cytC. Here, $\chi_{0s}$ is
    calculated for the entire simulation cell containing $N_s$
    waters. The solid lines is the fit of $\chi_{0s}(r) / \chi_{0s}$
    to an exponential function, $1- \exp[-(r-a)/\lambda]$, with 
    $\lambda= 24$ \AA\ and $a=2.6$ \AA. The dashed line connects the
    simulation points.  }
  \label{fig:4}
\end{figure}

Spatial correlation between the protein and water dipoles are found to
be long-ranged. This is illustrated in Fig.\ \ref{fig:4} where we plot
$\chi_{0s}(r)$ and $N(r)$ reduced to their values obtained from the
entire simulation box.  Function $N(r)$ is a number of waters in a
shell of thickness $r$ as measured from the protein van der Waals
(vdW) surface; $\chi_{0s}(r) \propto \langle \delta \mathbf{M}_0 \cdot
\delta\mathbf{M}_s(r)\rangle$ is the cross-correlation function
obtained from the total dipole moment $\mathbf{M}_s(r)$ of all waters
within the same shell.  Although $\chi_{0s}(r)$ clearly goes faster to
its saturation limit than the number of waters in the cell, it reaches
only half of its cell value for seven water shells around the
protein. In fact when $\chi_{0s}(r)/ \chi_{0s}$ is fitted to an
exponential decay function (solid line in Fig.\ \ref{fig:4}), the
corresponding correlation length turns out to be 24 \AA.

\begin{figure}
  \centering
  \includegraphics*[width=8cm]{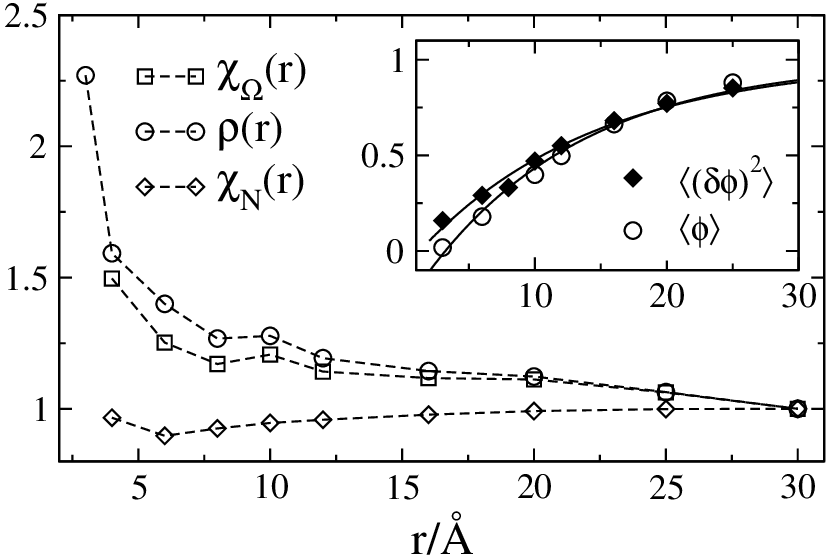}
  \caption{Dependence on the thickness $r$ of the solvation shell of
    the water number density $\rho(r)=N(r)/\Omega(r)$ and two dielectric
    susceptibilities, $\chi_{\Omega}(r)$ and $\chi_N(r)$ (see the text for
    the definitions). The inset shows the average potential $\langle
    \phi(r)\rangle$ at the Fe atom of the heme (open circles) and its
    variance $\langle (\delta\phi(r))^2\rangle$ (closed diamonds). The
    dashed lines in the main panel connect the points and the solid
    lines in the inset are fits of the simulation data to the
    function $1- \exp[-(r-a)/\lambda]$, with $\lambda= 12$ \AA\ and 13
    \AA\ for the potential and its variance, respectively. The
    calculations have been performed for cytC.  All results in the
    main panel are normalized to their corresponding values at $r=30$ \AA,
    while the data in the inset are normalized to the results obtained
    for the entire simulation box.  }
  \label{fig:5}
\end{figure}
 
The nearly disappearance of the cross-correlation between the protein
and shell dipoles at the shell thickness $r$ approaching the limit of
one hydration layer implies that waters in the first solvation shell
do not much correlate with the overall protein dipole and are more
driven by the local fields and vdW forces of the surface groups. It is
only more distant layers that correlate more extensively with the
global electrostatics of the solution. This picture contrasts with
a more localized density response of the interface shown in Fig.\
\ref{fig:5}. The density profile $\rho(r)=N(r)/\Omega(r)$ ($\Omega(r)$
is the shell volume) peaks at the interface, indicating wetting of the
protein surface by water, and then decays to approximately the bulk
density within $\sim 3$ solvation layers ($\simeq 10$ \AA).

We also show in Fig.\ \ref{fig:5} the dielectric susceptibility of the
surface waters determined by correlating the dipole moment of the
shell $\mathbf{M}_s(r)$ with the total dipole moment of water
$\mathbf{M}_s$.\cite{DMcpl:11} We define two susceptibilities,
normalized to the number of shell waters, $\chi_N(r) \propto N(r)^{-1}
\langle \delta\mathbf{M}_s(r)\cdot \delta\mathbf{M}_s\rangle$, and the
susceptibility normalized to the shell volume, $\chi_{\Omega}(r)
\propto \Omega(r)^{-1} \langle \delta\mathbf{M}_s(r)\cdot
\delta\mathbf{M}_s\rangle$. While the former is almost flat, showing
virtually no variation of polarity near the interface, susceptibility
$\chi_{\Omega}(r)$ shows about 30\% increase related to the
corresponding increase of the interfacial density. This polarity
increase is below the increment observed at the interface of a
non-polar solute with water,\cite{DMcpl:11} reflecting the topological
and chemical heterogeneity\cite{ChengRossky:98,Giovambattista:08} of
the protein surface.

Figure \ref{fig:5} suggests that interfacial properties scaling as
interfacial density will mostly decay to their bulk values within
$\sim 3$ hydration layers. However, this rule should not be blindly
extended to other observables. In particular, the electrostatic
response of the interface is only indirectly affected by density and
in principle can have a different range of convergence. This is shown
in the inset in Fig.\ \ref{fig:5} which presents the accumulation of
the average and variance of the electrostatic potential $\phi(r)$
produced by waters from the $r$-shell at the Fe atom of the heme. The
accumulation of both the average and the variance are much slower than
the decay of the interfacial density and are in fact comparable in
range to the accumulation of the cross-correlation $\chi_{0s}(r)$
shown in Fig.\ \ref{fig:4}. Specifically, when these data are fitted
to exponential decay functions, similarly to what has been done for
$\chi_{0s}(r)$, one gets the correlation lengths of 12--13 \AA.

\subsection{Dynamical properties}
\label{sec:4-2}
\begin{figure}
  \centering
  \includegraphics*[width=8cm]{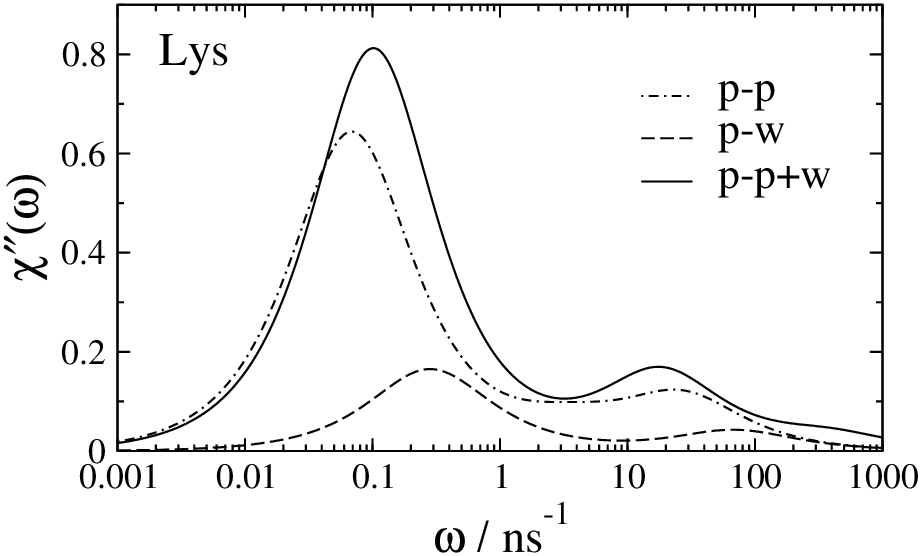}
  \caption{Normalized loss function ($\int_{0}^{\infty}
    \chi''(\omega)d\omega/(\pi\omega)=1$) obtained from the time
    cross-correlation function $S_0(t)$ (Eq.\ \eqref{eq:13}, solid
    line). The self protein-protein (dash-dotted line, p-p) and cross
    protein-water (dashed line, p-w) loss functions are obtained from the
    corresponding time correlation functions in Eq.\ 
    \eqref{eq:11}. Their relative weights reflect contributions of
    $\chi_{00}$ and $\chi_{0s}$ to $\chi_0^d$. The calculations were
    done for hydrated lys. }
  \label{fig:6}
\end{figure}

The long range of the dipolar protein-water correlation appearing in
susceptibility $\chi_{0s}$ will be masked in $\chi_0^d$ by a larger in
magnitude self protein component $\chi_{00}$. The two susceptibilities
exhibit, however, different dynamics, as is shown in Fig.\ \ref{fig:6}
for their loss functions. These are calculated from the time
correlation function
\begin{equation}
  \label{eq:13}
  S_{0}(t)= \left[ \langle \delta
    \mathbf{M}_0\cdot\delta\mathbf{M}\rangle \right]^{-1} \langle \delta
    \mathbf{M}_0(t)\cdot\delta\mathbf{M}(0)\rangle 
\end{equation}
and the corresponding self and cross correlation
functions
\begin{equation}
  \label{eq:11}
  S_{0a}(t)= \left[ \langle \delta
    \mathbf{M}_0\cdot\delta\mathbf{M}_a\rangle \right]^{-1} \langle \delta
    \mathbf{M}_0(t)\cdot\delta\mathbf{M}_a(0)\rangle  ,
\end{equation}
where $a=0,s$.  In the case of lys, there is about a factor of four
difference between the main Debye peaks of $\chi_{00}(\omega)$
($\simeq 14$ ns) and $\chi_{0s}(\omega)$ ($\simeq 3.5$ ns). There is
therefore a frequency window in which the two components of the
overall dipolar susceptibility can be separated. It is also worth
mentioning here that two relaxation times identified here for
$\chi_{0s}(\omega)$, 3.5 ns and 14 ps (lys), bracket the typical
dielectric $\delta$-relaxation band observed between protein and water
frequency peaks in dielectric loss functions of protein
solutions.\cite{Oleinikova:04} The solute-solvent dipolar
cross-correlations are considered as a plausible cause of the
$\delta$-dispersion.\cite{Boresch:00,Schroder:06,Rudas:06}

Both electrostatic and binding properties of surface waters are highly
heterogeneous, but this reality is differently reflected by the
observables. The dynamics of the protein dipole moment are essentially
single-exponential, decaying on the characteristic time of 4--14 ns of
protein tumbling.  The dynamics of the dipole moment of the first
hydration layer are also fairly generic. The correlation functions
decay much faster (see the SM\cite{supplJCP}), but still contain a
10--20\% slow component with the relaxation time close to that of the
protein dipole. This slow component can be assigned to waters strongly
attached to the protein surface.\cite{Bone:85}

\begin{figure}
  \centering
  \includegraphics*[width=7cm]{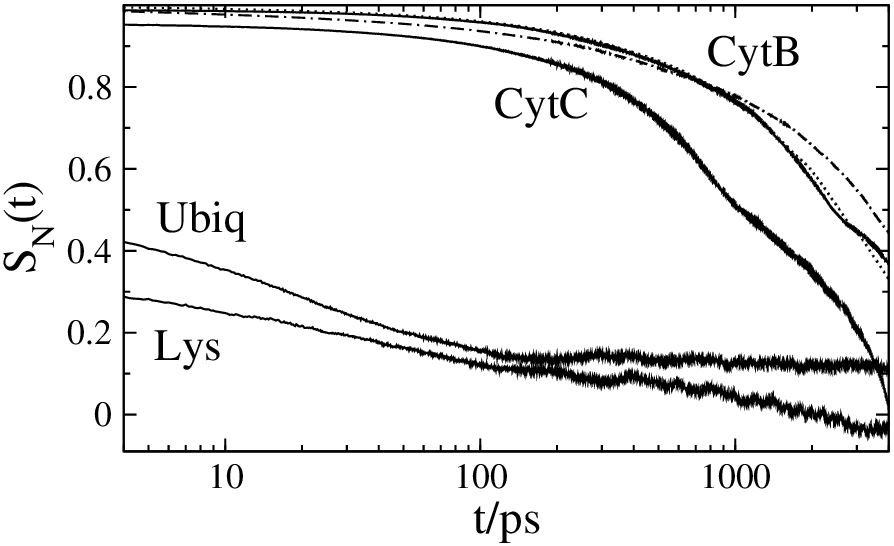} 
  \caption{Normalized time self-correlation functions of the number of
    waters in the first hydration shell $S_N(t)$.  The data are
    obtained from MD simulations; multi-exponential fits of the
    correlation functions are given in the SM.\cite{supplJCP} The
    dash-dotted and dotted lines indicate the self correlation
    functions $S_{00}(t)$ [Eq.\ \eqref{eq:11}] of the protein dipole
    for cytB and cytC, respectively. }
  \label{fig:7}
\end{figure}

In contrast to the generic dynamics of the protein and first-shell
dipoles, the population dynamics of the first layer are more specific.
The self-correlation functions $S_N(t)$ of the number of first-shell
waters $N_1(t)$ follows the rotational dynamics of $\mathbf{M}_0(t)$
for cytC and cytB (Fig.\ \ref{fig:7}). It appears that waters in the
first layer of these two proteins are strongly bound to the protein
surface positionally, but can rotate relatively freely, resulting in
fast relaxation of the first-shell dipole moment. On the contrary,
waters are weakly bound to the surface of lys and ubiq, with
significantly faster decays of their populations (Fig.\ \ref{fig:7}).

The binding affinity of the first-shell waters is, however,
heterogeneous. This is seen particularly clear for ubiq. Its initial
relaxation is two-exponential, with relaxation times of 0.2 ps (58\%)
and 39 ps (30\%). This initial fast decay is followed, however, by a
plateau indicating that about 12\% of first-shell waters do not leave
the protein surface on the simulation time-scale. A similar long-time
component, contributing to about 17\% of the correlation function, was
found for cytB (see the SM\cite{supplJCP}). 

In addition to the differences in the first-shell population dynamics
seen for the two pairs of proteins, the first-shell compressibilities
$\langle(\delta N_1)^2 \rangle/ \langle N_1\rangle$ are quite
different for them as well (last column in Table \ref{tab:1}). The
compressibilities of first shells of lys and ubiq are similar to those
found for water shells around rigid non-polar
solutes,\cite{MittalHummer:08,DMcpl:11} while they are much higher for
the two cytochromes. It is not yet clear how this observation relates
to the corresponding differences in the population dynamics.

To connect these observations to our electrostatic problem, it seems
clear that different observables variably report on the protein-water
interfacial structure. The dipolar response of a protein is
particularly sensitive to the distribution of the surface charge and
might potentially be considered as a marker of a given protein.  High
variability of $\epsilon_0$ among the proteins offers a potential for
applications to protein detection and separation in solution.

\section{Dielectrophoresis of protein solutions}
\label{sec:5}
Inhomogeneous external electric field exerts a force on the solute
dipole.\cite{JonesBook:95,HughesBook:03} This is given by the expression
\begin{equation}
  \label{eq:6}
  \mathbf{F}_0 = \frac{3\Omega_0}{8\pi\epsilon_s} K \nabla E_0^2 ,
\end{equation}
where $K=(4\pi/3)\epsilon_s(\chi_0^d + \chi_0^{\text{int}})$ is the
dielectrophoresis constant.  The solute is dragged toward a stronger
field if $K>0$ (positive dielectrophoresis) or toward a weaker field
if $K<0$ (negative dielectrophoresis).

We will use Eq.\ (\ref{eq:21}) to remove dipolar cross-correlations
and recast $K$ in terms of susceptibilities $\chi_{00}$ and
$\chi_1$. One gets
\begin{equation}
  \label{eq:7}
   K  = \frac{\epsilon_s +2}{3} y_0   +\frac{4\pi}{3}\epsilon_s\chi_1\left[1+
     \frac{2}{3} (\epsilon_s-1) y_0 \right],  
\end{equation}
where the dipolar density of the solute $y_0 = y_e +
(4\pi/3)\chi_{00}$ is introduced in analogy to the dipolar density of
a polar liquid\cite{Scaife:98,SPH:81} $y=(4\pi/9)\beta \rho m^2$;
here, $m$ and $\rho$ are the dipole moment and number density of a
liquid. We have included the component $y_e$ into $y_0$ originating
from electronic polarizability of the solute, which has not been
considered so far, but needs to be included in high-frequency
calculations. This component is additive to the one arising from the
permanent dipole\cite{SPH:81} and can be connected to the measurable
refractive index of the protein $n_0$ by the Clausius-Mossotti
equation, $y_e = (n_0^2-1)/(n_0^2+2)$.

The two limiting, Lorentz and Maxwell, scenarios are worth emphasizing
here. In the Lorentz scenario, $\chi_1=0$ and one gets
\begin{equation}
  \label{eq:45}
    K_{\text{L}} = \frac{\epsilon_s+2}{3} y_0 . 
\end{equation}
The factor in front of $y_0$ can be recognized as the Lorentz cavity
field correction.\cite{Scaife:98} Further, this equation allows only
positive dielectrophoresis. 

In the Maxwell scenario, $\chi_1=\chi_1^{\text{M}}$ [Eq.\ (\ref{eq:24})], one gets
\begin{equation}
  \label{eq:46}
  K_{\text{M}} =  \frac{3\epsilon_s}{2\epsilon_s+1} y_0 - \frac{\epsilon_s -
    1}{2\epsilon_s+1} .
\end{equation}
Now, the factor in front of $y_0$ is the standard cavity field of
traditional dielectric theories.\cite{Scaife:98} Negative
dielectrophoresis is now allowed when the threshold value $y_0^* =
(\epsilon_s-1)/(3\epsilon_s)$ is reached.

The standard analysis of dielectrophoresis of non-polar colloidal
suspensions \cite{HughesBook:03} neglects the dipole moment of the solute,
assuming $y_0\simeq 0$ in Eq.\ \eqref{eq:46}. In addition, when a
dielectric constant $\epsilon_0$ can be assigned to the material of
the colloidal particle, $\epsilon_s$ in the second summand of Eq.\ 
\eqref{eq:46} is replaced by $\epsilon_s/ \epsilon_0$, according to the
standard rules for the electrostatics of dielectric
interfaces.\cite{Landau8} The result is the traditional constant of
dielectrophoresis commonly used in the analysis of colloidal
suspensions\cite{WangPethig:93}
\begin{equation}
  \label{eq:8}
  K = \frac{\epsilon_0 -\epsilon_s}{\epsilon_0+2\epsilon_s} .
\end{equation}
This approximation is not applicable to protein solutions and we will
instead explicitly consider the dipolar response of the protein, with
susceptibilities $\chi_{00}$ and $\chi_1$ in Eq.\ (\ref{eq:7}) from
MD simulations.

Dielectrophoresis measurements are typically performed with
oscillatory external fields. The static properties considered so far
need to be replaced with frequency-dependent response function
according to the standard rules of handling time correlation 
functions.\cite{Hansen:03} While the frequency-dependent dielectric
constant of water is well defined and tabulated from laboratory
measurements, more care is required to define the frequency dependent 
susceptibility $\chi_1(\omega)$ and $y_0(\omega)$. 

The standard rules of connecting the response functions to time
correlation functions sugest the following form for the
frequency-dependent response functions of the solute dipole
\cite{DMpre:10}
\begin{equation}
  \label{eq:47}
  \chi_{0a}(\omega) = \chi_{0a}\left[1 - i\omega \tilde S_{0a}(-\omega) \right],
\end{equation}
where $a=0,s$ and $\tilde S_{0a}(\omega)$ is the Fourier-Laplace
transform of the corresponding time correlation functions in Eq.\
(\ref{eq:11}). These two relations, with $\tilde S_{0a}(\omega)$
obtained from MD simulations, can be used in Eq.\ (\ref{eq:21}) to
find $\chi_1(\omega)$. In doing that, one needs to replace
$\epsilon_s$ with the frequency-dependent, complex-valued dielectric
constant of water $\epsilon_s(\omega)$. Once this is done, one can
employ Eq.\ (\ref{eq:7}) with the frequency-dependent susceptibilities
and $\mathrm{Re}[K(\omega)]$ in Eq.\ (\ref{eq:6}) to calculate the
force.\cite{WangPethig:93,HughesBook:03} The effect of the electrolyte
can be included into the water dielectric constant
$\epsilon_s(\omega)=\epsilon_D(\omega) + \sigma_i/(i\omega)$ through
the ionic conductivity of the solution $\sigma_i$, in addition to the
Debye dielectric constant $\epsilon_D(\omega)=\epsilon_D'(\omega) + i
\epsilon_D''(\omega)$ for the dielectric response of pure water.

\begin{figure}
  \centering
  \includegraphics*[width=8cm]{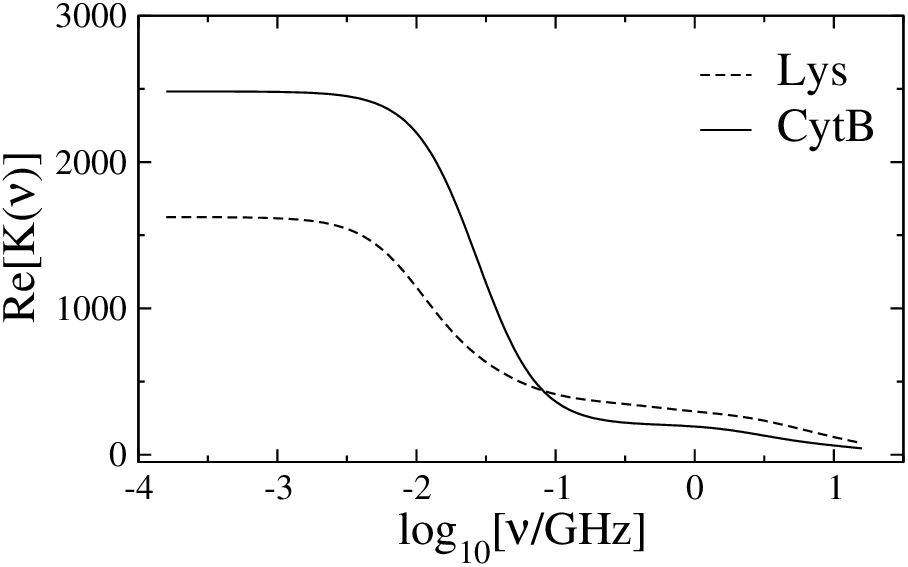}
  \caption{$\mathrm{Re}[K(\nu)]$, $\omega=2\pi\nu$ from Eq.\
    \eqref{eq:7} with the frequency-dependent response functions in
    Eq.\ (\ref{eq:47}) from MD simulations and $\epsilon_D(\omega)$ of
    water from Ref.\ \onlinecite{Yada:09}; the refractive index of the
    protein $n_0=1.47$.  }
  \label{fig:8}
\end{figure}

The results of calculations of $\mathrm{Re}[K(\omega)]$ for cytB and
lys are shown in Fig.\ \ref{fig:8}. The dynamics of the protein's
dipole, $\chi_{00}(\omega)$ and $\chi_{0s}(\omega)$ are taken directly
from MD simulations (see the SM\cite{supplJCP} for the fits of the
relaxation functions and the list of the relaxation times). The
frequency-dependent dielectric constant of water $\epsilon_D(\omega)$
is taken from Ref.\ \onlinecite{Yada:09} and $n_0=1.47$ is assigned to the protein. The
electrolyte conductivity in the range $\sigma_i\simeq 1-100$
mSm$^{-1}$ typically employed in the laboratory measurements
\cite{HughesBook:03} does not affect the outcome of the calculations.

The main result of these calculations is a dominance of $y_0(\omega)$,
arising from the protein permanent dipole, in the overall
dielectrophoresis response.  This result holds for all charged
proteins studied here.  In contrast, even static dielectrophoresis
constant is negative for ubiq, suggesting negative dielectrophoresis
in the entire frequency range.  Numerical difficulties in obtaining
$S_{0s}(t)$ for ubiq have prevented us from presenting $K(\omega)$ for
this protein.

\section{THz absorption of protein solutions}
\label{sec:6}
\begin{figure}
  \centering
  \includegraphics*[width=8cm]{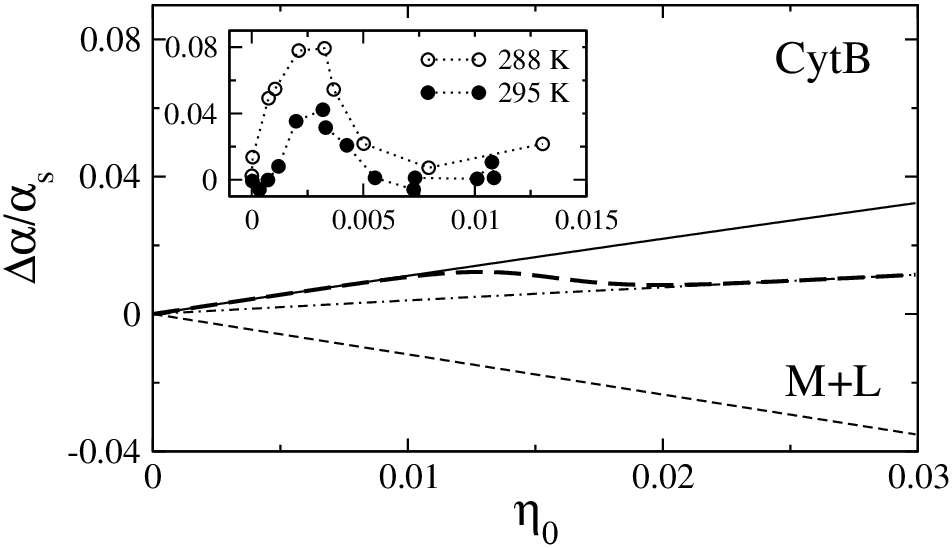}
  \caption{Absorbance increment calculated at the frequency of 2.25
    THz vs the molar fraction $\eta_0$ of cytB in water. The
    calculation with $\chi_1(\omega)$ from MD simulations (solid line)
    is compared to $\chi_1(\omega)$ from either Maxwell or Lorentz
    scenarios (dashed line, labeled ``M+L'') which give very close
    results in this frequency range. The dash-dotted line is the
    calculation performed with the restriction
    $(4\pi/3)\epsilon_s(\omega)\chi_1(\omega)=1$ corresponding to zero
    $\chi_{0s}$ in Eq.\ \eqref{eq:21}. The bold dashed line is
    arbitrary drawn to illustrate a crossover between two linear
    slopes when proteins in solution start to affect each other and
    prevent full convergence of $\chi_{0s}(r)$ characterized by a long
    correlation length of $\simeq 2.4$ nm [Fig.\ \ref{fig:4}]. The
    inset shows experimental data reported in Ref.\
    \onlinecite{Ebbinghaus:07} for a five-helix bundle protein
    $\lambda^*_{6-85}$ at two temperatures indicated in the plot. The
    dotted lines in the inset connect the points. }
  \label{fig:9}
\end{figure}

Positive $\chi_1$, instead of the negative values in both the Lorentz
and Maxwell scenarios, alter the theory predictions concerning the
absorption of THz radiation by protein solutions. We have recently
suggested a model,\cite{DMpre:10} based on the formalism of response
functions discussed in Sec.\ \ref{sec:3}, to address this problem.
The study was motivated by experimental reports of the positive slope
of the absorbance increment with an increasing concentration of the
protein in solution, following by a non-linear downward turn of the
concentration dependence\cite{Ebbinghaus:07,Born:09} (inset in Fig.\
\ref{fig:9}).  Such a trend, recorded in the 1--3 THz frequency range,
contradicts the traditional view that adding a protein, less polar
than water, should lower the solution polarity and the corresponding
radiation absorbance.  Indeed, the model calculations in Ref.\
\onlinecite{DMpre:10} could not account for the observations without
demanding a significant increase of the effective dipole of the
protein-water interface. That development was, however, based on the
Maxwell scenario for the interface susceptibility and thus negative
interface dipole.  Since the present formalism goes beyond the Maxwell
picture and the numerical results allow a positive interface dipole,
this problem needs to be revisited.

We are looking at the absorbance of the electromagnetic radiation\cite{Landau8} 
\begin{equation}
  \label{eq:55}
    \alpha(\omega) = \frac{4\pi\omega}{c}\
    \frac{\chi^{T''}(\omega)}{\sqrt{1+4\pi\chi^{T'}(\omega)} }, 
\end{equation}
where $c$ is the speed of light and the superscript ``T'' emphasizes
that we are considering the susceptibility of the solution
$\chi(\omega)$ in the direction of the electric field perpendicular
(transversal) to the direction of light propagation determined by the
wave-vector. The property reported experimentally is the relative
increment $\Delta \alpha(\omega) / \alpha_s(\omega)
=\alpha_{\text{mix}}(\omega) / \alpha_s(\omega) -1 $ of the solution
(mixture) absorbance over the absorbance of the homogeneous liquid
$\alpha_s(\omega)$.

The derivation of the solution absorbance directly follows from the
generalized response function in Eq.\ (\ref{eq:10}). The
derivation steps follow Ref.\ \onlinecite{DMpre:10} and are briefly
summarized in Appendix \ref{appB}. One gets the following result for
the relative susceptibility increment
\begin{equation}
  \label{eq:9}
   \begin{split}
  \Delta &\chi^T(\omega) / \chi^T_s(\omega) 
     = - \eta_0 \left[1 - \frac{4\pi}{3} \epsilon_s(\omega)\chi_1(\omega)
     \right]\\
    & +
    y_0(\omega)\eta_0\left[\frac{\epsilon_s(\omega)+2}{\epsilon_s(\omega)-1}
    +\frac{8\pi}{3}\chi_1(\omega)\epsilon_s(\omega)
    I_0(\eta_0,R) \right] ,
  \end{split}
\end{equation}
where $\Delta\chi^T(\omega)=\chi_{\text{mix}}^T(\omega)
-\chi^T_s(\omega)$ and $\chi_s^T(\omega)=
(\epsilon_s(\omega)-1)/(4\pi)$.  Further, $\eta_0$ is the volume
fraction of the solutes and $I_0(\eta_0,R)$ represents the mutual
polarization of the solutes by their permanent dipoles aligned by the
field of radiation, a non-ideal solution effect. It is given by Eq.\
\eqref{eq:57} in Appendix \ref{appB} and has been tabulated as a
function of $\eta_0$ and $R$ in Ref.\ \onlinecite{DMpre:10} assuming
hard-sphere structure factor for the solutes in solution. This latter
approximation has been used in the calculations presented in Fig.\
\ref{fig:9}.

All the parameters in Eq.\ (\ref{eq:9}) are defined in our
calculations of the dielectrophoresis response and are directly
applied to the calculations shown in Fig.\ \ref{fig:9} for cytB.  The
solid line shows $\Delta \alpha(\omega)/\alpha_s(\omega)$ calculated at
2.25 THz of radiation used in the experiment.\cite{Ebbinghaus:07} The
result is an obviously positive slope of the absorption increment,
which arises from both the interface and permanent dipoles
re-enforcing each other. In contrast, both Lorentz and Maxwell
scenarios suggest a negative slope (dashed line in Fig.\ \ref{fig:9}).

The non-linear dependence on the solute volume fraction present in
$I_0(R,\eta_0)$ in Eq.\ (\ref{eq:9}) turns out to be insufficient to
bend the concentration dependence downward (inset in Fig.\
\ref{fig:9}). Several possible long-range effects are still missing
from our analysis. The structure factor of solvated proteins is
estimated from its hard-sphere limit and can be modified by
long-ranged interactions. In addition, correlation between proteins'
dipole moments, represented by the Kirkwood factor, has not been
included in the present calculation.\cite{DMpre:10} We, however, want
discuss yet another possibility related to the long decay of the
solute-solvent dipolar correlations shown in Fig.\ \ref{fig:4}.

The correlation length of $\lambda \simeq 2.4$ nm found in Fig.\
\ref{fig:4} suggests that susceptibility $\chi_{0s}$ becomes affected
when proteins in solution come closer than $2\lambda$, which roughly
corresponds to the position of the peak in the experimental
concentration dependence in the inset of Fig.\ \ref{fig:9}. If mutual
effect of the proteins in solution does not allow $\chi_{0s}$ to
saturate, it implies that, according to Eq.\ (\ref{eq:21}), $\chi_1$
should change with the concentration and approach the limit
$(4\pi/3)\epsilon_s(\omega)\chi_1(\omega)\simeq 1$. This limit for the
absorbance is shown by the dash-dotted line in Fig.\ \ref{fig:9}. The
overall concentration dependence is expected to exhibit a crossover
from one linear slope to another in the concentration range where
proteins start to affect each other, as is schematically shown by the
bold dashed line in Fig.\ \ref{fig:9}. The experiment shows a
qualitatively similar crossover behavior. Further, no cross-over of
the concentration dependence of the absorption coefficient was found
for solutions of aminoacids.\cite{Niehues:2011fk} This observation
suggests a much shorter length-scale of dipolar cross-correlations for
hydrated aminoacids compared to proteins. The reason might be related
to a more homogeneous structure of waters around aminoacids and thus a
higher contribution of the first hydration layer to $\chi_{0s}$, which
is almost absent for proteins (Fig.\ \ref{fig:4}).

A brief comment on the experimental temperature dependence of the 
absorption shown in the inset of Fig.\ \ref{fig:9} is relevant here. 
The permanent dipole susceptibility $\chi_{00}$ is proportional to
$\beta$. The slope of the absorbance with concentration is therefore
expected to increase with lowering temperature, as is indeed
qualitatively observed.

These calculations and qualitative comparisons with experiment suggest
that most of the polar response of the protein charges is frozen on
the time-scale of THz radiation, which therefore allows one to probe
the polarization of the protein-water interface projected on the
dipole $M_0^{\text{int}}$. This interface polarization is distinctly
different from the Lorentz-Maxwell scenario and this fact is reflected
in the positive slope of the absorbance with increasing concentration.

\section{Summary}
\label{sec:7}
In conclusion, we have utilized long simulation trajectories and large
system sizes to systematically study the dipolar susceptibility [Eq.\
\eqref{eq:43}] of hydrated proteins. This property exhibits large
variation among the proteins, suggesting its possible use as a
``marker'' of a protein in solution. The large variance in the protein
susceptibility also opens the door to their detection and separation
in solution, in particular by dielectrophoresis of the protein
solutions.

The permanent dipole susceptibility is a combination of the variance
of the intrinsic protein dipole and a cross-correlation between the
protein and water dipoles.  The cross-correlation component is found
to be negative for uncharged ubiq, but is positive for the charged
proteins. It is also long-ranged, decaying into the bulk
on the correlation length of about 2 nm. A similar, but somewhat
smaller, decay length is found for the electrostatic potential
produced by the hydration water inside the protein.

Dipolar solute-water cross-correlations are significant and cannot be
neglected. In the case of ubiq the self and cross correlation
components nearly cancel each other, but produce a negative overall
susceptibility.  This protein therefore demonstrates negative
dielectrophoresis and, correspondingly, a dia-electric dipolar
response.

Thermodynamic stability prohibits dia-electric response ($ \epsilon_s
< 1$) for bulk dielectrics.\cite{Landau8} This limitation does not
apply to a finite-length polar response,\cite{Dolgov:81} and in fact
the wave-vector dependent dielectric constant $\epsilon_s(k)$ is
negative in a certain range of $k$-values for most polar
liquids.\cite{Raineri:99} Thermodynamic arguments also do not rule out
a dia-electric response of solutes in solution. Negative
dielectrophoresis is in fact fairly common for colloidal
suspensions.\cite{HughesBook:03} Nevertheless, this is the first
report of negative dielectrophoresis of proteins by numerical
simulations.

An important question posed by the present study and requiring further
investigation is whether negative values of $\chi_0$ 
found here for ubiq are general for neutral proteins.  If the answer
to this question is affirmative, dielectrophoresis of hydrated
proteins is expected to be sensitive to the buffer pH (as indeed has
been reported \cite{HughesBook:03}). In addition, negative
dielectrophoresis, and thus dia-electric response, should be common
when the buffer pH approaches protein's isoelectric point. Future
simulations and laboratory measurements are required to shed more
light on this intriguing perspective. In terms of physiological
conditions of protein activity, this possibility would imply a range
of pH values in which a protein is nearly insensitive to significant
gradients of local electric fields, i.e., ``invisible'' to local
fields.

The interface polarization is a small portion of the static or
low-frequency dipolar response, but gains in importance at high
frequencies, above the protein Debye peak representing its rotation in
solution. In this frequency range, the polar response of the protein
charges is dynamically arrested and the response of much faster
interfacial waters starts to show up. Here, different dielectric
scenarios (such as Lorentz or Maxwell recipes) predict distinctly
different outputs. Numerical simulations presented here show a broad
range of possibilities, beyond those emphasized by these two
electrostatic limits, depending on the distribution of the protein
surface charge. These different outcomes are reflected by the THz
absorbance of the protein solution. In particular, the sign of the
slope of the THz absorbance vs the protein concentration reports on
whether the interface dipole is oriented along the polarizing field or
opposite to it.  A parallel orientation corresponds to a positive
slope, whereas an anti-parallel orientation (such as in the Maxwell
scenario) yields a negative slope. The measurements of the THz
absorbance give therefore a direct access to the polarization pattern
of the protein-water interface.

\acknowledgments This research was supported by the National Science
Foundation (CHE-0910905). CPU time was provided by the National
Science Foundation through TeraGrid resources (TG-MCB080116N). Useful
discussions with Alexandra Ros are gratefully acknowledged. David
LeBard has kindly shared his code (Pretty Fast Analysis) for 
analyzing protein trajectories with GPU acceleration. 

\appendix
\section{Derivation of Eq.\ (\ref{eq:21})}
\label{appA}
The function $\bm{\chi}(\mathbf{k}_1,\mathbf{k}_2)$, first calculated
in Refs.\ \onlinecite{DMjcp1:04} and \onlinecite{DMjcp2:04}, describes
the response of a polar liquid outside a void.  The latter is defined
by the step function $\tilde\theta_0(\mathbf{k})$ [Eq.\
(\ref{eq:20})]. This function and its conjugate $\tilde\theta = \tilde
1 - \tilde\theta_0$, $\tilde 1(\mathbf{k}_1,\mathbf{k}_2) = (2\pi)^3
\delta(\mathbf{k}_1-\mathbf{k}_2)$ together form a set of orthogonal
functions projecting the polar response and corresponding electric
fields inside and outside the solute. They obey the following
orthogonality and multiplication rules: $\tilde\theta*\tilde \theta_0
=0$, $\tilde\theta * \tilde \theta = \tilde \theta$, $\tilde\theta_0 *
\tilde \theta_0 = \tilde \theta_0$, where, as above, asterisks refers
to the $\mathbf{k}$-space convolution.

The inhomogeneous dipolar response function can then be written as
follows
\begin{equation}
  \label{eq:26}
  \bm{\chi} = \bm{\chi}_s - \alpha \bm{\chi}_s*\tilde\theta_0 *
  \mathbf{G}^{-1} * \tilde\theta_0*\bm{\chi}_s .
\end{equation}
Here, $ \bm{\chi}_s(\mathbf{k}_1,\mathbf{k}_2) =
\bm{\chi}_s(\mathbf{k}_1)\tilde 1(\mathbf{k}_1,\mathbf{k}_2)$ and
$\mathbf{G}=\tilde\theta_0*\bm{\chi}_s*\tilde\theta_0$, which in terms
of explicit wave-vectors implies
\begin{equation}
  \label{eq:48}
  \mathbf{G}(\mathbf{k}_1,\mathbf{k}_2) =
  \tilde\theta_0(\mathbf{k}_1-\mathbf{k}')*\bm{\chi}_s(\mathbf{k}',\mathbf{k}'')*
  \tilde\theta_0(\mathbf{k}'',\mathbf{k}_2). 
\end{equation}

Equation (\ref{eq:26}) with $\alpha=1$ presents an exact solution of
the problem of dipolar response of a polar liquid interfacing a
spherical void.\cite{DMjcp1:04,DMjcp2:04} It does not, however,
anticipate the formation of the specific orientational structure at
the interface characteristic of liquid
water.\cite{Lee:84,Sokhan:97,Bratko:09,Rossky:10} Therefore,
$\alpha=\chi_1/ \chi_1^{\text{M}} \ne 1$ is introduced in Eq.\
(\ref{eq:26}) to account for possible deviations from the Maxwell
scenario of the surface polarization.

From the definition of the response function, the projection of the
polarization inside the solute is
\begin{equation}
  \label{eq:27}
  \tilde\theta_0*\mathbf{\tilde P} =
  (1-\alpha)\tilde\theta_0*\bm{\chi}_s*\mathbf{\tilde E}_0. 
\end{equation}
The polarization projection is of course zero in case of $\alpha=1$
since this is how the response function was
constructed.\cite{DMjcp1:04} A modification of the inhomogeneous
response by $\alpha\ne 1$ in the second term in Eq.\ (\ref{eq:26}) is
physically equivalent to creating a non-zero dipole associated with
the solute. In case of a uniform external field $\mathbf{\tilde
  E}_{0}(\mathbf{k}) =\delta_{\mathbf{0},\mathbf{k}} \mathbf{\hat
  z}E_{0}$ this dipole becomes
\begin{equation}
  \label{eq:49}
  M_{0z} = \mathbf{\hat z}\cdot\tilde\theta_0(-\mathbf{k})*\mathbf{\tilde P}(\mathbf{k}) =
  (1-\alpha) \chi_s^L(0) \Omega_0 E_0 . 
\end{equation}

When the electric field is produced by the solute dipole [Eq.\
(\ref{eq:17})], this field can be written as the projection of the
dipolar field outside the solute excluded volume $\mathbf{\tilde E}_0
= \tilde\theta*\mathbf{\tilde E}_d$. Here, $\mathbf{\tilde E}_d$ is
the Fourier-space electric field of a point dipole, without the space
cutoff equating the field to zero inside the solute and responsible
for the appearance of the spherical Bessel function in Eq.\
(\ref{eq:18}). With this representation one gets from Eq.\
(\ref{eq:27})
\begin{equation}
  \label{eq:28}
  \tilde\theta_0*\mathbf{\tilde P} =
  (1-\alpha)\tilde\theta_0*\bm{\chi}_s*\tilde\theta*\mathbf{\tilde E}_d.
\end{equation}
If $\bm{\chi}_s(\mathbf{k})$ here is replaced with its
continuum limit $\bm{\chi}_s(0)$, the integral becomes identically zero
because of the orthogonality of $\tilde\theta_0$ and $\tilde\theta$. 
The result is $\mathbf{M}_s=\mathbf{\tilde P}(0)$ in Eq.\
(\ref{eq:23}). 

If we now choose the direction of $z$-axis along $\mathbf{M}_0$ and
take the projection of the solute dipole on $\mathbf{\hat z}$, we get 
\begin{equation}
  \label{eq:30}
  M_{s,z}/M_0=\mathbf{\hat z} \cdot
  \bm{\chi}(0,\mathbf{k})*\mathbf{\tilde T}(\mathbf{k})\cdot
  \mathbf{\hat z} .     
\end{equation}
A note on how to correctly take the $k\rightarrow 0$ limit in the
$\mathbf{k}$-space tensor functions is appropriate here. Since both
$\mathbf{\hat z}$ and $\mathbf{\hat k}$ impose axial symmetry on the
liquid, which is isotropic in the direct space, one has to take
$\mathbf{\hat k}$ parallel to $\mathbf{\hat z}$ when taking the
$k\rightarrow 0$ limit to avoid imposing a bi-axial symmetry. In this
prescription, $\tilde T_{zz}(0) = - (8\pi/3)$ and one gets for the
homogeneous component of $\bm{\chi}(0,\mathbf{k})$ (first summand in
Eq.\ \eqref{eq:26})
\begin{equation}
  \label{eq:31}
  \mathbf{\hat z} \cdot \bm{\chi}_s(0) \cdot \mathbf{\tilde T}(0) \cdot
  \mathbf{\hat z} = -\frac{2(\epsilon_s-1)}{3\epsilon_s} .
\end{equation}

Similarly, one can write down the inhomogeneous component of the polar
response from Eq.\ (\ref{eq:10}) as
\begin{equation}
  \label{eq:32}
   \begin{split}
 (\chi_1^{\text{M}})^{-1}  \mathbf{\hat z} \cdot &\bm{\chi}_0(0,\mathbf{k}) *\mathbf{\tilde T}(\mathbf{k}) \cdot
  \mathbf{\hat z} =\\
                  &\frac{6\Omega_0 }{\pi R^2\chi_s^L(0)} \int_0^{\infty} dk j_1(kR)^2 
  \langle\mathbf{\hat z} \cdot
  \bm{\chi}_s(\mathbf{k})\cdot\mathbf{\tilde D}\cdot\mathbf{\hat
    z}\rangle_{\omega_k}  , 
  \end{split}
\end{equation}
where $\langle\dots\rangle_{\omega_k}$ refers to the average over the
solid angles of the unit vector $\mathbf{\hat k}$ and
\begin{equation}
  \label{eq:33}
  \tilde\theta_0(k)/ \Omega_0 = 3 \frac{j_1(kR)}{kR}
\end{equation}
has been used. Using the relations $\langle \mathbf{\hat z}
\cdot\mathbf{J}^L \cdot \mathbf{\hat z}\rangle_{\omega_k}=(1/3)$ and  $\langle \mathbf{\hat z}
\cdot\mathbf{J}^T \cdot \mathbf{\hat z}\rangle_{\omega_k}=(2/3)$ one gets in the continuum
limit $k\rightarrow 0$ for the response function
$\bm{\chi}_s(\mathbf{k})$
\begin{equation}
  \label{eq:34}
 (\chi_1^{\text{M}})^{-1}  \mathbf{\hat z} \cdot \bm{\chi}_0(0,\mathbf{k}) *\mathbf{\tilde T}(\mathbf{k}) \cdot
  \mathbf{\hat z} = -(8\pi/9) (\epsilon_s - 1), 
\end{equation}
where 
\begin{equation}
  \label{eq:35}
  (6R/ \pi) \int_0^{\infty} j_1(kR)^2 dk = 1
\end{equation}
and Eqs.\ (\ref{eq:29}) for the $k=0$ values of $\chi_s^{L,T}(0)$ have
been used. Combining Eqs.\ (\ref{eq:31}) and (\ref{eq:34}), one
arrives at Eq.\ (\ref{eq:21}). 

A similar procedure can be applied to calculate the dipole moment of
the solvent in uniform external field $\mathbf{\tilde
  E}_{0}(\mathbf{k})
=\delta_{\mathbf{0},\mathbf{k}} \mathbf{\hat z}E_{0}$. The
dipole moment in this case is
\begin{equation}
  \label{eq:36}
  M_{s,z} = \mathbf{\hat z}\cdot
  \tilde\theta(-\mathbf{k}')*\bm{\chi}(\mathbf{k}',\mathbf{k})*
  \mathbf{\tilde  E}_{0}(\mathbf{k}) .
\end{equation}
According to the preceding arguments,
$\tilde\theta*\bm{\chi}=\bm{\chi} -
(1-\alpha)\tilde\theta_0*\bm{\chi}_s $. From this relation one gets
\begin{equation}
  \label{eq:37}
  M_{s,z} = M_{\text{liq}} + M_{0z}^{c} = M_{\text{liq}} - M_{0z} + M_{0z}^{\text{int}},
\end{equation}
where $M_{0z}$ is given by Eq.\ (\ref{eq:49}) and $M_{\text{liq}}$ is
the dipole induced by $E_{0}$ in a homogeneous liquid without the
solute. Further, $M_{0z}^{c}$ in Eq.\ \eqref{eq:37} is the ``cavity
dipole'' associated with inserting the excluded solute volume into the
polarized liquid. It is given as
\begin{equation}
  \label{eq:51}
  M_{0z}^c/(\Omega_0 E_0) = \chi_1 - \chi_s^L(0)\left(1-\chi_1/
  \chi_1^{\text{M}} \right).
\end{equation}
Equations (\ref{eq:37}) and (\ref{eq:51}) yield Eq.\ (\ref{eq:50}) for
the overall solution dipole. 

\section{Derivation of Eq.\ \eqref{eq:9}}
\label{appB}
We want to calculate the dipole moment $M^T(\omega)$ of the solution
induced by a polarized electromagnetic wave propagating along the
$z$-axis of the laboratory frame and having the electric field vector
along the $x$-axis of the same frame. The wave-vector is therefore
along $\mathbf{\hat z}$ and the response is transversal. The
susceptibility in Eq.\ \eqref{eq:55} becomes
\begin{equation}
  \label{eq:16}
  \chi^T(\omega) = M^T(\omega)/(VE_0) .
\end{equation}

The formalism of response functions\cite{DMjcp1:04,DMpre:10} described in
Appendix \ref{appA} gives the following prescription for the
calculation of the dipole moment of the solution projected on $\mathbf{\hat x}$
\begin{equation}
  \label{eq:19}
  M^T = \chi_{00}\Omega_0 N_0 E_0 +\mathbf{\hat x}\cdot
  \tilde\theta(\mathbf{k})*\bm{\chi}(\mathbf{k},\mathbf{k}')*\mathbf{\tilde
    E}_0(\mathbf{k}') .
\end{equation}
The first term in this equation (dependencies on frequency are omitted
for brevity) is the direct polarization of the permanent dipoles of
$N_0$ solutes in the mixture and the second term is the polarization
of the solvent. 

The formalism of Appendix \ref{appA} is now extended to an
ensemble of solutes. This extension is straightforward if polarization
fields of the liquid around different solutes are uncorrelated.\cite{DMpre:10}
From Eqs.\ \eqref{eq:10} and \eqref{eq:26} one gets
\begin{equation}
  \label{eq:22}
  \bm{\chi}(\mathbf{k}_1,\mathbf{k}_2)=
  \bm{\chi}_s(\mathbf{k}_1)\delta_{\mathbf{k}_1,\mathbf{k}_2} - \alpha
  \sum_i 
   \bm{\chi}_0(\mathbf{k}_1,\mathbf{k}_2)
   e^{i\mathbf{k}_2\cdot\mathbf{r}_i}  ,
\end{equation}
where the sum runs over the $N_0$ solutes in the solution. In
addition, the source of the electric field now includes the
homogeneous external field and the field of all dipolar solutes
polarized by it
\begin{equation}
  \label{eq:38}
  \mathbf{\tilde E}_0(\mathbf{k}) = E_0\delta_{\mathbf{0},\mathbf{k}}
  \mathbf{\hat x} +  \chi_{00}\Omega_0 E_0 \sum_i \mathbf{\tilde
    T}(\mathbf{k})\cdot\mathbf{\hat x}
  e^{-i\mathbf{k}\cdot\mathbf{r}_i} .
\end{equation}

According to the rules of calculating the $\mathbf{k}$-space
convolutions explained in Appendix \ref{appA}, the substitution of the
uniform external field (first summand in Eq.\ \eqref{eq:38}) into Eq.\
\eqref{eq:19} leads to
\begin{equation}
  \label{eq:42}
  \frac{M^T_1}{VE_0} = \chi_s^T(0)\left[ 1 - \eta_0 
    \left(1 + \alpha \frac{\epsilon_s-1}{2\epsilon_s+1}\right)\right],
\end{equation}
where $\eta_0=\Omega_0N_0/ V$ is the solute volume fraction. 

Similarly, the use of the second, dipolar summand from Eq.\
\eqref{eq:38} in Eq.\ \eqref{eq:19} yields 
\begin{equation}
  \label{eq:56}
  \frac{M^T_2}{VE_0} = \frac{4\pi}{3}\eta_0\chi_s^T(0) \chi_{00} 
   \left[1 - \alpha I_0(\eta_0,R)
     \frac{2(\epsilon_s-1)}{2\epsilon_s+1} \right] . 
\end{equation}
Here, one employs $\tilde T_{xx}(0)=(4\pi)/3$ for the transverse
component of the dipolar tensor and 
\begin{equation}
  \label{eq:57}
  I_0(\eta_0,R) = \frac{6R}{\pi}\int_0^{\infty} j_1(kR)^2 S_0(k) dk,
\end{equation}
where
\begin{equation}
  \label{eq:58}
  S_0(k) = N_0^{-1} \sum_{i,j} e^{i\mathbf{k}\cdot(\mathbf{r}_i - \mathbf{r}_j)}
\end{equation}
is the density structure factor\cite{Hansen:03} of the solutes in
the solution. 

The integral $I_0(\eta_0,R)$ is equal to unity for an ideal solution
with $S_0(k)=1$ [as in the case of Eq.\ \eqref{eq:35}].  It is also
tabulated in Ref.\ \onlinecite{DMpre:10} in terms of a polynomial
interpolation in the solute size and volume fraction in the
hard-sphere approximation for $S_0(k)$.  This latter approximation was
used in the calculations shown in Fig.\ \ref{fig:9}. Finally,
$M^T=\chi_{00}\Omega_0N_0 E_0+ M^T_1+M^T_2$ from Eqs.\ \eqref{eq:42}
and \eqref{eq:56} yields the susceptibility increment in Eq.\
\eqref{eq:9}.

We note that $y_e$ from protein polarizability has been added to
$(4\pi/3)\chi_{00}$ in Eq.\ \eqref{eq:9}. One can use experimental
refractive index of the protein powder in the same frequency range as
for the solution absorption to estimate the contribution of the
protein intrinsic dipole from the Clausius-Mossotti equation
$(n_0(\omega)^2-1)/(n_0(\omega)^2 +2) = y_e + (4\pi/3)\chi_{00}$ the
use of which should be restricted to THz and higher frequencies.


%

\end{document}